\documentclass[10pt,letterpaper]{article}
\usepackage{opex3}
\usepackage{subfigure}
\usepackage{cite}
\usepackage{ae} %%for Computer Modern fonts
\usepackage[normalem]{ulem}

\begin{document}

%%%%%%%%%%%%%%%%%% title page information %%%%%%%%%%%%%%%%%%
\title{Integrated CARS source based on seeded four-wave mixing in silicon nitride}

\author{J\"{o}rn P. Epping$^{1,*}$, Michael Kues$^{2}$, Peter J.M. van der Slot$^{1}$, Chris J. Lee$^{1,3,4}$, Carsten Fallnich$^{2}$ and Klaus-J. Boller$^{1}$}

\address{$^{1}$Laser Physics \& Nonlinear Optics Group, Faculty of Science and Technology, MESA$^{+}$ Research Institute for Nanotechnology, University of Twente, P. O. Box 217, Enschede 7500AE, The Netherlands
 			\\$^{2}$Institut f\"ur Angewandte Physik, Westf\"alische Wilhelms-Universit\"at M\"unster, Corrensstra\ss e 2, 48149 M\"unster, Germany
			\\$^{3}$FOM Institute DIFFER, Edisonbaan 14, 3439 MN Nieuwegein, The Netherlands
			\\$^{4}$XUV Optics Group, Faculty of Science and Technology, MESA$^{+}$ Research Institute for Nanotechnology, University of Twente, P. O. Box 217, Enschede 7500AE, The Netherlands}

\email{$^{*}$j.p.epping@utwente.nl} %% email address is required



\begin{abstract}
We present a theoretical investigation of an integrated nonlinear light source for coherent anti-Stokes Raman scattering (CARS) based on silicon nitride waveguides. Wavelength tunable and temporally synchronized signal and idler pulses are obtained by using seeded four-wave mixing. We find that the calculated input pump power needed for nonlinear wavelength generation is more than one order of magnitude lower than in previously reported approaches based on optical fibers. The tuning range of the wavelength conversion was calculated to be 1418~nm to 1518~nm (idler) and 788~nm to 857~nm (signal), which corresponds to a coverage of vibrational transitions from 2350~cm$^{-1}$ to 2810~cm$^{-1}$. A maximum conversion efficiency of 19.1\% at a peak pump power of 300~W is predicted.
\end{abstract}

 % REPLACE WITH CORRECT OCIS CODES FOR YOUR ARTICLE

%%%%%%%%%%%%%%%%%%%%%%% References %%%%%%%%%%%%%%%%%%%%%%%%%

%%%%%%%%%%%%%%%%%%%%%%%%%%  body  %%%%%%%%%%%%%%%%%%%%%%%%%%
\section{Introduction}

Coherent anti-Stokes Raman scattering (CARS) offers the ability to probe both the vibrational frequencies and coherences of materials, making it attractive for many applications. This includes monitoring the temperature and reaction dynamics in combustion engineering~\cite{hall1981}, remote sensing of explosives~\cite{Portnov2008}, stand-off infectious agent detection~\cite{Beadie2005}, as well as chemically selective imaging~\cite{evans2008}. The latter application is particularly challenging since full hyperspectral imaging requires rather specialized light sources. Indeed, the majority of CARS imaging systems aim for only one or two vibrational features, due to the limitations of the sources and signal-to-noise considerations.

The CARS process itself is based on a partially resonant four-wave mixing (FWM) process with at least two input frequencies, whose frequency difference encompasses one or more vibrational transitions.
Spectral resolution can then be achieved by a number of means, such as using a broadband excitation pulse and a narrow bandwidth probe pulse~\cite{Kee2004}, temporal delay between frequency chirped excitation and probe~\cite{Rocha2008}, and through the use of tunable, narrow bandwidth excitation pulses~\cite{jurna2006}. The first method relies on light pulses with a broad spectral bandwidth, and selectivity is provided mainly by the spectral bandwidth of the probe pulse. The second method also relies on femtosecond light pulses stretched to a few ps by applying a linear chirp to the spectrum, and the spectral resolution relies on giving the pump and probe equal chirp. The use of tunable narrow bandwidth excitation pulses, on the other hand, allows for better signal-to-noise ratios. This comes, however, at the expense of the complexity of the light sources used to provide excitation and probe pulses. In order to achieve spectral resolution at a relatively high peak power, at least two temporally synchronized pulses in the picosecond regime are preferred, because their spectral bandwidth is in the order of the bandwidth of typical vibrational transitions~\cite{evans2008}. Furthermore, the pulses have to be widely tunable, such that their difference frequency covers different vibrational transitions, which lie typically below 3200~cm$^{-1}$.

Often complex light sources are used, such as optical parametric oscillators~\cite{jurna2006} or electronically synchronized laser oscillators~\cite{potma2002}. Recently, all-fiber approaches have been introduced, which take advantage of the wide natural bandwidth of FWM in optical fibers---especially photonic crystal fibers, pumped near the zero dispersion wavelength. The optical pulse bandwidth can be reduced through filtering~\cite{lamb2013}, temporal focusing~\cite{Rocha2008}, and injection seeding~\cite{lefrancois2012, baumgartl2012, gottschall2012}. 
A major advantage of all-fiber approaches, compared to lasers and optical parametric oscillators, is their potential compactness and relatively easy maintenance. Another advantage is that temporal synchronization can be achieved by carefully designing the fiber dispersions without the need of an external delay stage~\cite{baumgartl2012}. Injection seeding is an attractive option, because it allows control over the emitted signal and idler spectra and increases the spectral power densities needed for CARS applications. Nevertheless, fiber solutions lack compatibility with integrated photonics. This makes it difficult to couple CARS light sources with microfluidic devices for biological sensing applications~\cite{Ymeti2005}, and integrated light sources~\cite{Oldenbeuving2012,saha2013} for ease of tunability. 

A source based on integrated photonics is of great interest for analyzing CARS spectra in a lab-on-a-chip setup, which may provide label-free analytical techniques~\cite{Camp2009}. It is known that efficient wavelength conversion in integrated devices is possible due to high index contrast and using materials with a higher nonlinear response compared to silica based fibers. Indeed, efficient FWM has been shown in highly nonlinear waveguide materials like silicon~\cite{foster2006}, chalcogenide glasses~\cite{luan2009} and doped silica~\cite{ferrera2008}. More recently silicon nitride (Si$_3$N$_4$), which is also widely used in lab-on-a-chip applications, has been applied for waveguide-based FWM. Silicon nitride waveguides with core thickness beyond 500~nm have become available~\cite{levy2010, agha2012, luke2013}. These increased core thicknesses are sufficient to place the zero dispersion wavelength (ZDW) in the near-infrared, allowing efficient phase-matched FWM. Silicon nitride has a relatively high Kerr nonlinearity~\cite{ikeda2008} and an appropriate transparency range for the near IR wavelengths typically used for CARS. Unlike semiconductor or chalcogenide waveguides, two-photon losses are reduced because the bandgap is relatively large (5~eV)~\cite{Sze}. Finally, linear power losses can be extremely low ($\sim$1~dB/cm)~\cite{bauters2011}, allowing for relatively long interaction lengths.

Here, we show, through a theoretical investigation, that a silicon nitride waveguide-based CARS light source has many attractive properties. In the normal dispersion region the signal and idler gain spectra become widely spaced, providing a frequency difference between pump and idler that is well-suited to probe the vibrational spectrum of condensed matter samples. By injecting an additional narrow-band, continuous wave seed, with a wavelength that overlaps the signal gain spectrum, a longer-wavelength idler pulse is generated with a narrow bandwidth, instead of the broad bandwidth usually obtained through spontaneous FWM. Furthermore, the required pump laser peak power is found to be one order of magnitude less than that required for fiber based approaches.

\section{Integrated CARS source}

We consider a stoichiometric silicon nitride ridge waveguide with a rectangular cross-section deposited on a silica substrate. In order to have maximum conversion efficiency for a frequency difference between pump (taken as 1064~nm) and idler in the range below 3200~cm$^-1$, we choose the zero dispersion wavelength to be at 1069 nm. The corresponding waveguide dimensions are then a width of 1630~nm and a height of 700~nm. Different pump wavelengths can also be accommodated by using different waveguide dimensions, however, all the following calculations are for the given pump wavelength of 1064~nm as it is readily available in practice. In order to numerically study the nonlinear dynamics of FWM in waveguiding structures, a time-dependent calculation of the light field in ideally three spatial dimensions would be required, however, this is computationally challenging. Instead, we use the one-dimensional generalized nonlinear Schr\"odinger equation~\cite{agrawal}. To reduce the problem to one spatial dimension, the nonlinear coefficient $\gamma = n_2 \omega /(cA_{eff})$~\cite{agrawal} and the dispersion of the waveguide is pre-calculated by solving for the transverse mode profiles of the fundamental modes for the pump, signal and idler fields with a finite-element mode solver. In the expression for the nonlinear coefficient $\gamma$, $n_2$ is the nonlinear refractive index , $\omega$ the pump frequency, A$_{eff}$ the effective area of the fundamental quasi-TE mode and $c$ the speed of light. From the mode profile, the nonlinear coefficient was calculated to be $1.93$~m$^{-1}$W$^{-1}$, where $n_2 = 2.4\times10^{-15}$~cm$^{2}$/W, is taken from \cite{ikeda2008}, Ikeda et al.

\begin{figure}[htbp]
	\centering
	\mbox{
		\subfigure{
			\includegraphics[height=5cm]{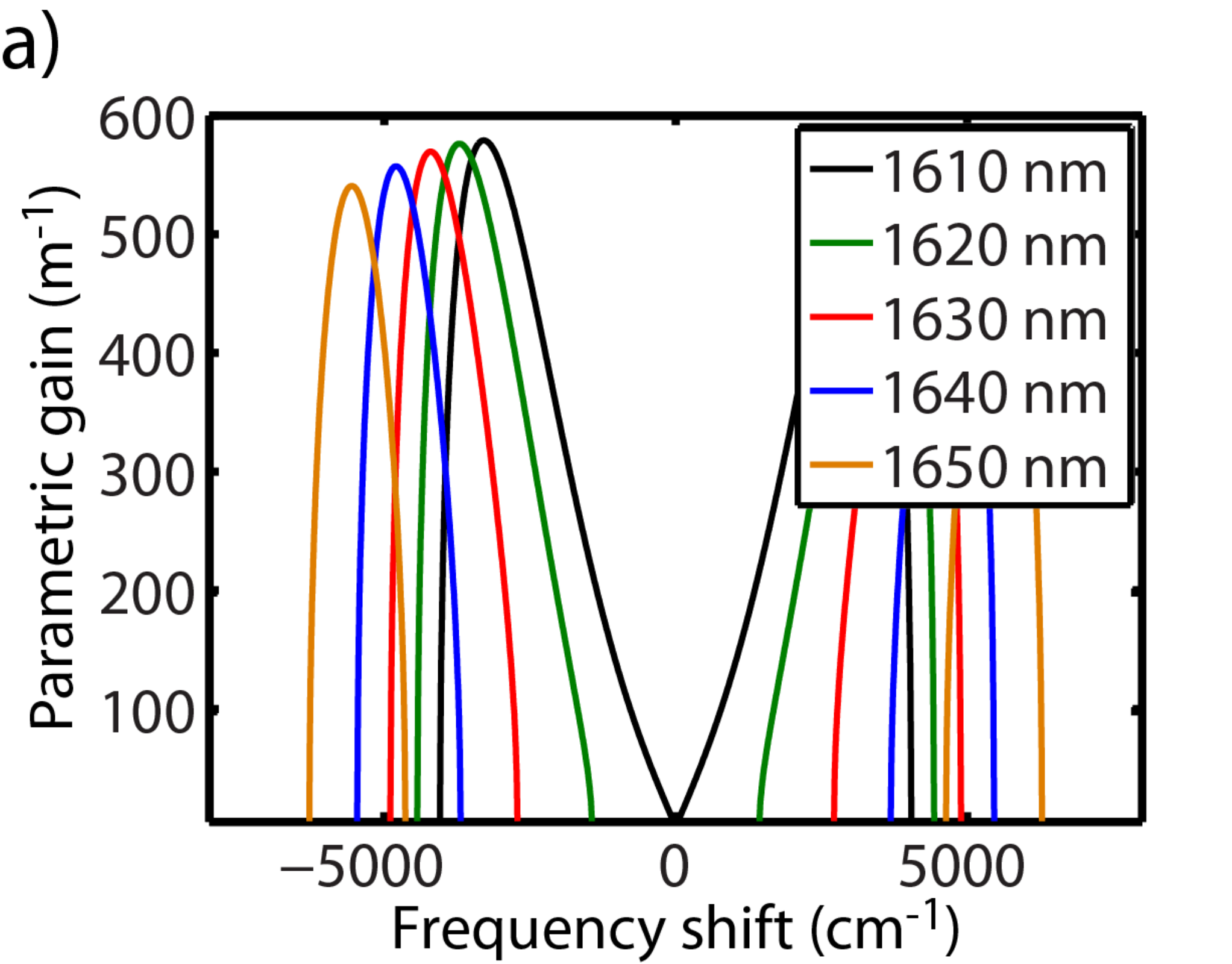}}
				\quad
		\subfigure{
			\includegraphics[height=5cm]{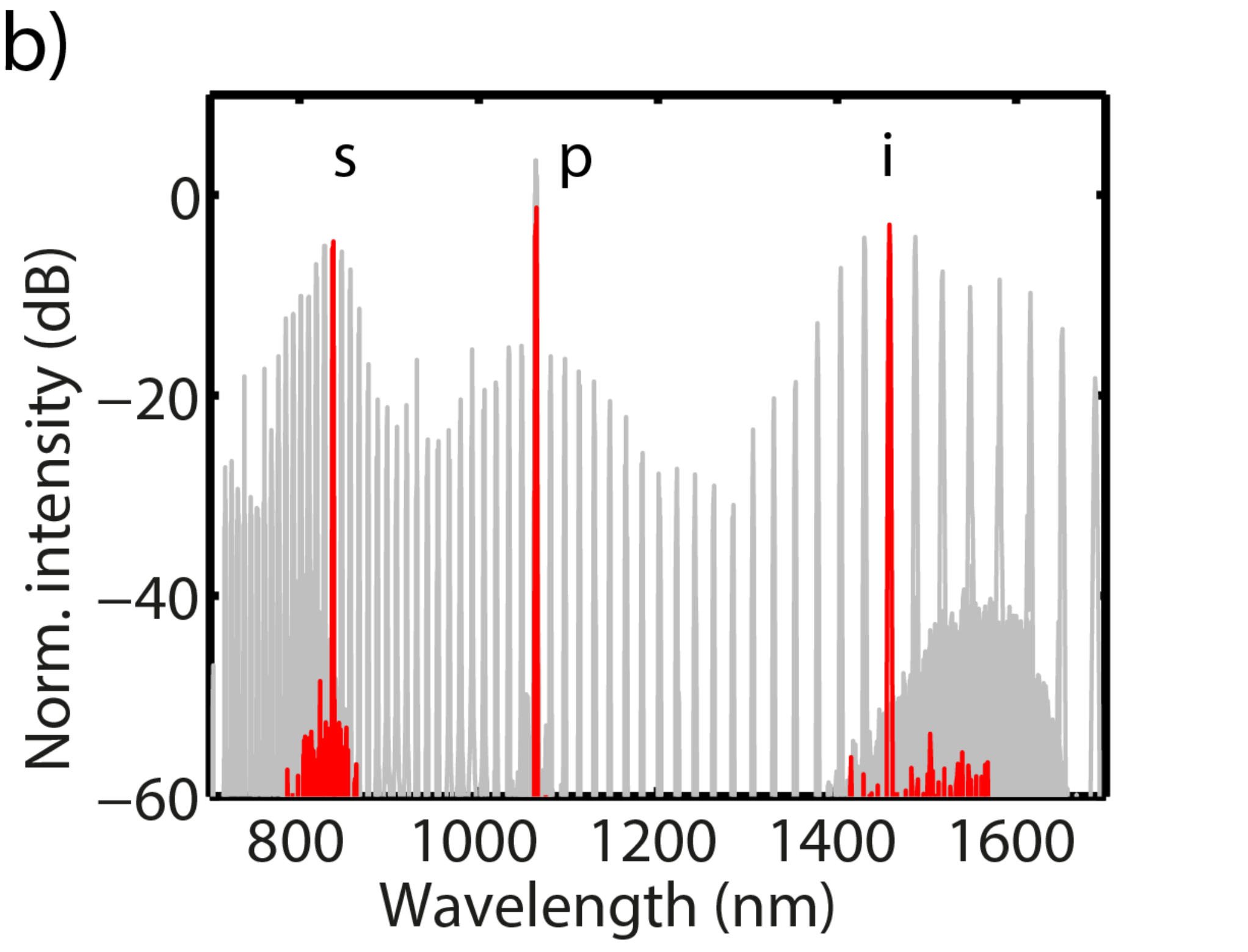}}
	}
	\caption{	\label{fig:1}(a) Analytically calculated FWM small-signal gain spectra of silicon nitride waveguides with a height of 700 nm and widths from 1610 nm to 1650 nm for a pump power of 300~W. (b) Superimposed spectra of cw seeded FWM after 2~cm of propagation for a pump pulse with 300~W peak power, which was numerically calculated using the nonlinear Schr\"odinger equation. Shown in red is a single spectrum, obtained with a seed wavelength (s) of 828~nm, and resulting in an idler wavelength (i) of 1488~nm, and pumped (p) at 1064~nm. The spectra for a range of seed wavelengths from 714~nm to 1063~nm in 4~THz steps are shown in grey.}
\end{figure}

The spectrum of the FWM small-signal gain ~\cite{agrawal} for the fundamental quasi-TE mode was calculated using $\gamma$ and the waveguide dispersion relations obtained from the finite element calculations.
The results for waveguide widths from 1610~nm to 1650~nm are displayed in Fig.~\ref{fig:1}(a), where a pump wavelength, $\lambda_{p}$, of 1064~nm and a power, $P_0$, of 300~W have been assumed. The full width at half maximum (FWHM) of the FWM gain spectrum, as in Fig. \ref{fig:1}(a), is 1735~cm$^{-1}$ at a waveguide width of 1630~nm (red curve). This significant bandwidth is sufficient to scan over a large range of vibrational levels. Furthermore, Fig.~\ref{fig:1}(a) shows that the peak in the parametric gain can be shifted to a different frequency by changing the waveguide width. This is due to the change in location of the zero dispersion wavelength, and the resulting change in waveguide dispersion determines the region where the FWM process is phase matched. The vibrational frequency spectrum of interest lies at frequencies below 3200~cm$^{-1}$. The spectral coverage of the FWM in Fig.~\ref{fig:1}(a) is mostly at a higher frequency difference. However, the full nonlinear calculations presented below show that the depletion of the pump results in smaller frequency shifts, which are in the desired range.

The seeded FWM pulse propagation is calculated by numerically integrating the generalized nonlinear Schr\"odinger equation using a split-step Fourier method as reported in \cite{kues2009}, Kues et al. The slowly varying envelope approximation has been applied to the light field, that includes shot noise~\cite{genty2007}. 
To obtain a spectrally narrow CARS Stokes pulse, the idler spectrum is set by injecting a single-frequency, continuous wave at the signal frequency, $\omega_{s}$, with a power, $P_s$, of 100~mW. 
The temporal shape of the pump pulse is Gaussian with a FWHM of 10~ps. The peak powers are varied between 150~W and 350~W, which is in the range of waveguide-based lasers \cite{xiao2008}. These parameters result in idler pulses with a bandwidth that is narrower than that typical for vibrational transitions, which are in the range from 10 to 20~cm$^{-1}$~\cite{evans2008}, but still offer the peak powers needed for CARS at feasible average powers.

The propagation of the pulses through the waveguide takes into account nine orders of dispersion from the material and waveguide. The linear power loss of the waveguide is taken as 1~dB/cm over the whole wavelength range, as was reported for a silicon nitride ridge waveguide with similar dimensions~\cite{agha2012}. The Raman effect is neglected compared to FWM. This is justified for amorphous glasses, as is considered here~\cite{agrawal}. 

\begin{figure}[htbp]
	\centering
	\mbox{
		\subfigure{
					\includegraphics[height=5cm]{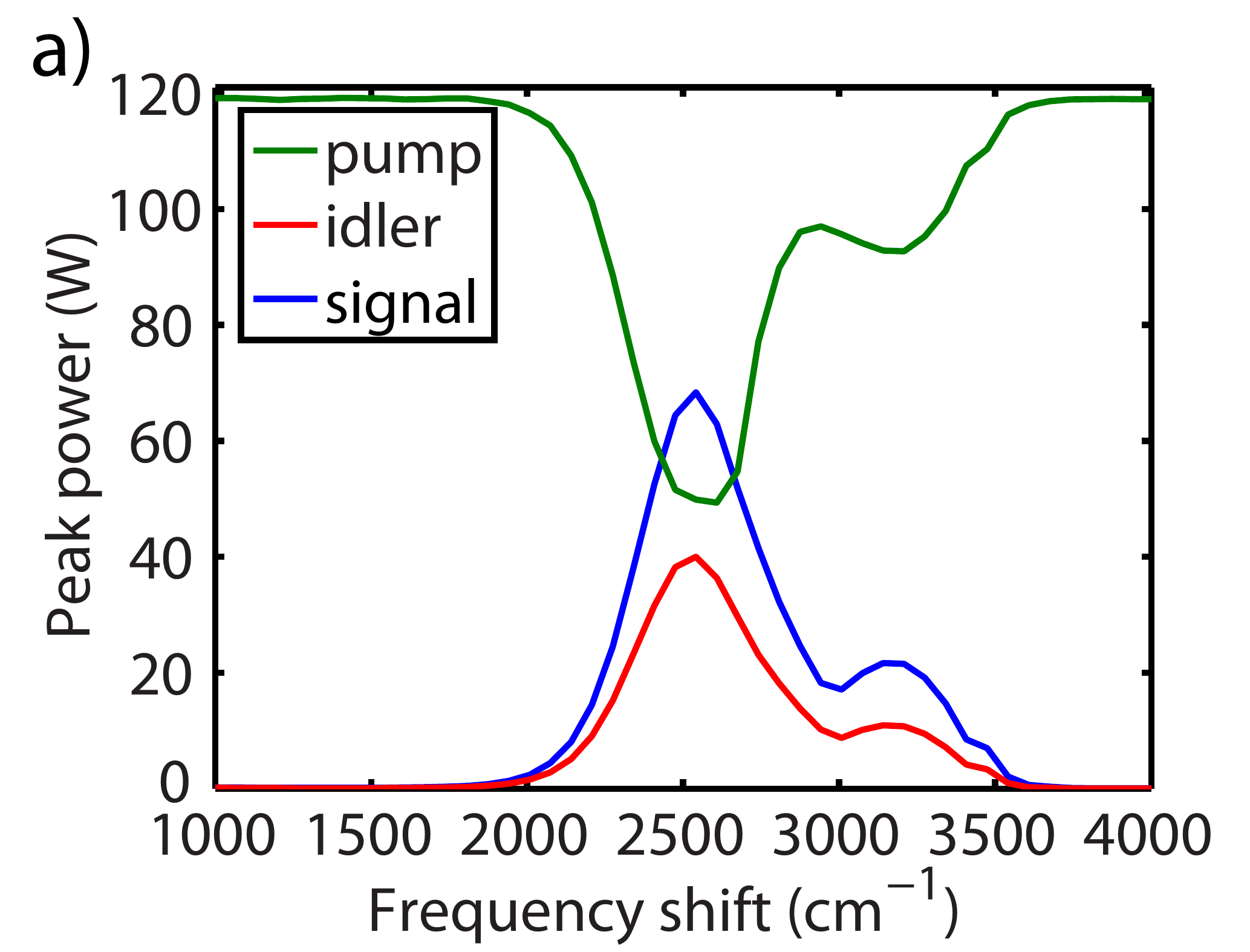}}
				\quad
		\subfigure{
					\includegraphics[height=5cm]{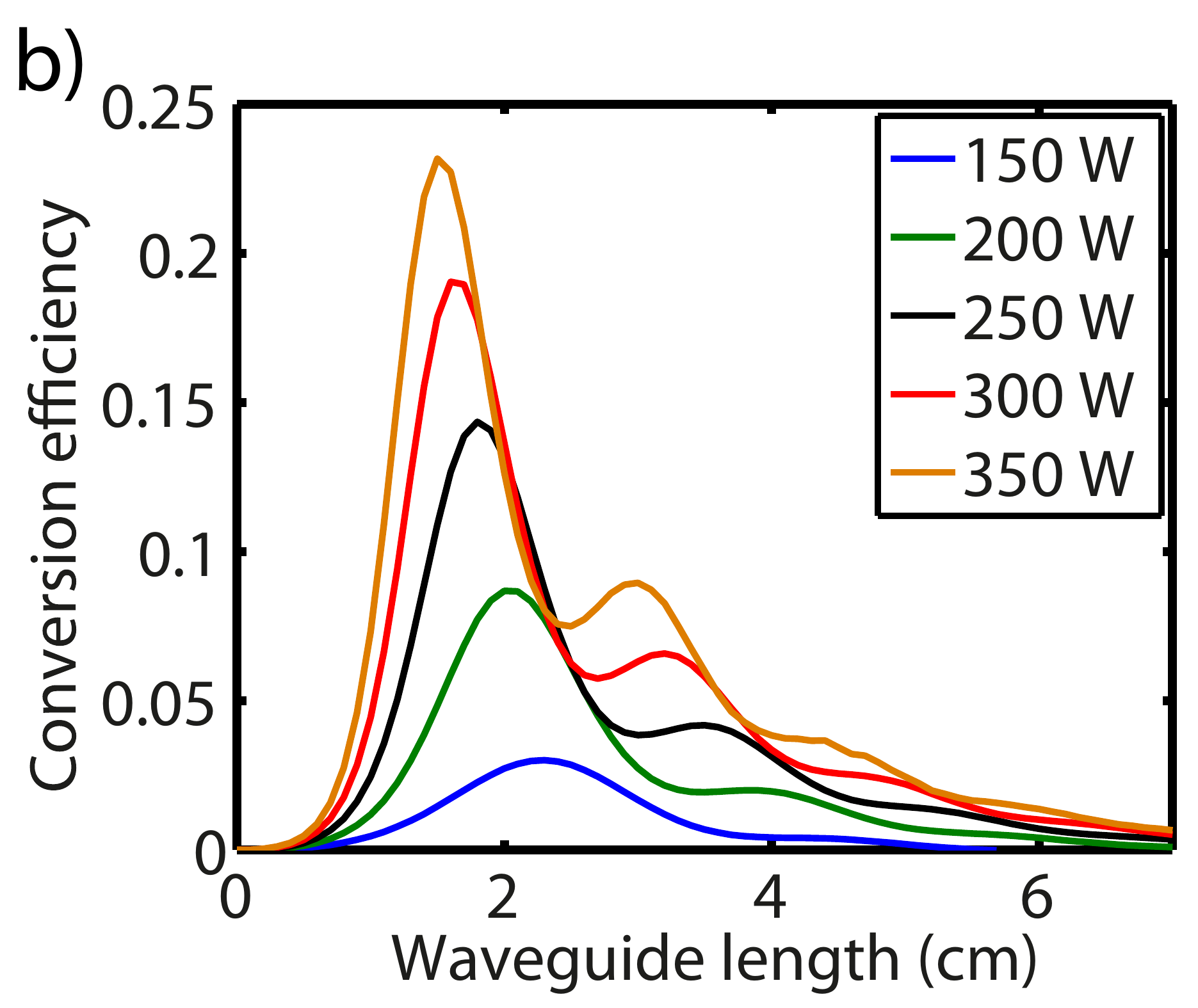}}
	}
		\caption{\label{fig:3} (a) Resulting peak powers of pump, signal and idler pulse against the frequency shift from the pump frequency. These results were calculated for a waveguide with a width of 1630~nm, a height of 700~nm, and length of 2~cm. The peak pump power was set to be 300~W. (b) Calculated conversion efficiencies against propagation for peak pump powers from 150~W to 350~W seeded at a wavelength of 828~nm. }
\end{figure}

In order to show that the injection seeding imposes narrowband signal and idler spectral output tunable over the gain bandwidth, rather than the spontaneous broadband FWM spectrum, the output spectrum was calculated for seed wavelengths in the range of 714--1063~nm in 4~THz steps, a peak pump power of 300~W and a waveguide width of 1630~nm as can be seen in Fig.~\ref{fig:1}(b). It can be seen that FWM occurs over the entire simulated range, however, we require that at least 20~W of peak power are available for CARS. Using this criteria, it can be seen that both the signal and idler are narrowband, and idler (signal) are generated with the required strength in the wavelength range from 1418 to 1518~nm (851 down to 819~nm). The difference frequency between the pump and idler corresponds to vibrational frequencies in the range of 2346~cm$^{-1}$ to 2810~cm$^{-1}$. Comparing these results with Fig.~\ref{fig:1}(a), we observe that the tuning range of 464~cm$^{-1}$ is not only smaller than the one observed in Fig.~\ref{fig:1}(a) but the range has moved closer to the pump frequency as well. Clearly, the small-signal gain approach overestimates the tuning range and absolute shift of signal and idler frequency with respect to the pump, due to the fact that the calculation of the small-signal gain assumes an undepleted pump~\cite{agrawal}, while the nonlinear Schr\"odinger equation includes the full nonlinear dynamics. The peak power of the pump, signal and idler is shown in Fig.~\ref{fig:3}(a) as a function of the frequency shift. This figure shows that maximum peak powers of 68.4~W and 40.0~W are obtained for the signal and idler, while the maximum pump depletion results in a minimum peak pump power of 49.4~W. 

Moreover, over the tuning range of 2346~cm$^{-1}$ to 2810~cm$^{-1}$, the peak power of the idler pulses remains above 20~W, which is large enough to give a strong CARS signal. The calculated seeded idler intensities in this tuning range are at least 25 to 30~dB higher than the noise level of competing spontaneous FWM (which is broadband). This low background will not contribute significantly to the generated CARS signals.

We calculated the conversion efficiency, defined as $\eta = (E_s + E_i -E_{s,0})/(E_{p,0} + E_{s,0})$, where $E_{p,0}$ is the injected pump energy, and $E_{s,0}$=2~pJ is the energy injected by the signal seed over the time period of the simulation window (20~ps). $E_s$ and $E_i$ are the pulse energies of the generated signal and idler pulses. The results are shown in Fig. \ref{fig:3}(b) for five different peak pump powers from 150~W to 350~W (with $\lambda_p$=1064 nm, $\tau_p$=10~ps and $P_s$ = 100~mW) as a function of waveguide length. The seed wavelength is held constant at 828~nm, which corresponds to an idler wavelength of 1488~nm and matches the peak of the FWM gain. For pump peak powers of 100~W or less, no significant FWM was observed.

For peak powers of 150~W and above the conversion efficiency at first shows strong exponential growth until a maximum is reached. The maximum is due to propagation losses and pump depletion, both of which lower the overall FWM output. After reaching a maximum back conversion starts to set in. For a peak power of 300~W, a maximum conversion efficiency of 19.1~\% is reached after 1.8~cm of propagation. The calculated pulse length of the generated idler pulses is 6.3~ps after a propagation of 1.8~cm with a spectral width (FWHM) of 1.2~nm, which is about two times more than the Fourier limit. The corresponding spectral bandwidth of 5~cm$^{-1}$ is still smaller than the bandwidths of typical vibrational transitions.

A maximum conversion of 23.2~\% was calculated after 1.5~cm for a peak power of 350~W and back conversion is observed for longer interaction lengths. This shows that, in order to achieve a maximum conversion efficiency, the length of the waveguide section where FWM takes place has to be carefully selected using calculations that include nonlinear dynamics. Note that the FWM process can easily be stopped by quickly changing the local dispersion of the waveguide, which destroys the phase matching between the waves, by tapering the waveguide, for example. 

\begin{figure}[htbp]
	\centering\includegraphics[width=7cm]{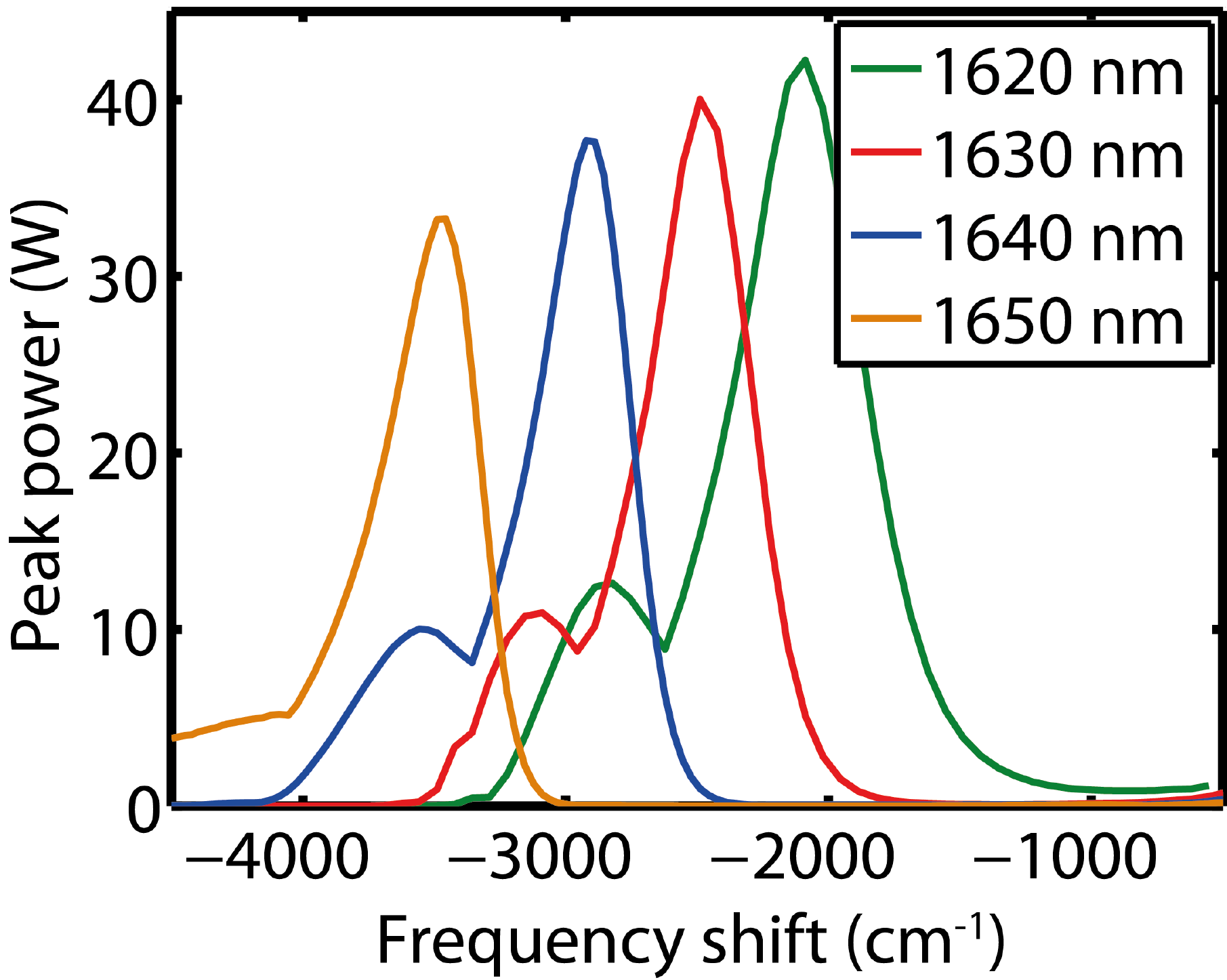}
		\caption{\label{fig:4}Peak power of the idler pulses against the frequency shift for various waveguide widths. The waveguide height is 700~nm, and the interaction length is 2~cm, while the pump peak power is 300~W.}
\end{figure}

The desired FWM gain spectrum can be controlled by carefully selecting the waveguide dispersion. In order to get to the appropriate dispersion the most obvious parameter to change is the waveguide width since it can be controlled precisely during fabrication of the integrated waveguides. In Fig. \ref{fig:4} the calculated peak powers of idler pulses is shown as a function of the frequency shift for waveguide widths ranging from 1610~nm to 1650~nm and a height of 700~nm, while pumped with the same parameters as in Fig. \ref{fig:3}(b). The maximum idler peak power range from 42.3~W at 2088~cm$^{-1}$ ($\lambda_{i}$ = 1368~nm) at a waveguide width of 1620~nm to 33.3~W at 3456~cm$^{-1}$ ($\lambda_{i}$ = 1683~nm) at a width of 1650~nm. The lower conversion efficiency for broader waveguides is explained by the larger effective mode area, $A_{eff}$, which results in a lower nonlinear coefficient $\gamma$. When comparing the idler peak power of Fig. \ref{fig:4} with the small-signal gain of Fig. \ref{fig:1}(a), it can be seen that the full nonlinear Schr\"odinger equation predicts the maximum in the idler peak power to occur at a smaller difference frequency than the location of the maximum in the small-signal FWM gain. This can be explained by the power dependence of the FWM gain, since the peak power changes during propagation, due to temporal broadening of the pump pulse, propagation losses, and the high nonlinear conversion efficiency in silicon nitride waveguides. This highlights the necessity for numerical studies to predict the exact FWM gain spectrum.

\section{Conclusion}
In conclusion, we have shown, through numerical calculations, an efficient way to realise synchronized and tunable picosecond pulses with properties that are highly suitable for application in CARS microscopy and spectroscopy. A high conversion efficiency of 19.1~\% is calculated assuming a relatively high power loss of 1~dB/cm and a moderate pump peak power of 300~W, which is one order of magnitude lower than in previous reported fiber based approaches. The calculated peak powers as well as the wavelengths of the pump and idler pulse are in the range required for CARS experiments with picosecond pulses. The tuning range of our approach can easily be adjusted by changing the pump wavelength or the waveguide dimensions, because the bandwidth of the FWM gain strongly depends on the dispersion of the waveguides. Furthermore, the spectral coverage can even be doubled by using the signal in place of the pump in a CARS microscope or spectrometer (so the signal becomes the pump for the CARS process), because the signal and idler are of comparable peak power with the pump. These results show that an integrated CARS light source with dimensions on the length scale of 2~cm can be realized with silicon nitride waveguides at moderate laser powers that are available from waveguide based pulsed and continuous wave laser sources. This approach can easily be adapted for different wavelength ranges and pulse durations as well as for other waveguide platforms.

\section*{Acknowledgments}
This research was funded by the Stichting Technische Wetenschappen under the Generic Technologies for Integrated Photonics perspectief program (grant number 11358).

\end{document}